\documentclass{article}

\usepackage{microtype}
\usepackage{algorithm2e}
\usepackage{amsmath}
\usepackage{graphicx}
\usepackage{color}

\title{Annotating Synapses in Large EM Datasets}

\author{Stephen M. Plaza, Toufiq Parag, Gary B. Huang, Donald J. Olbris, \\
Mathew A. Saunders, and Patricia K. Rivlin}

\begin{document}
\maketitle

\begin{abstract}
Reconstructing neuronal circuits at the level of synapses is a central
problem in neuroscience and becoming a focus of the emerging field of 
connectomics. To date, electron microscopy (EM) is the most proven technique 
for identifying and quantifying synaptic connections.  As advances in EM make 
acquiring
larger datasets possible, subsequent manual synapse identification ({\em i.e.}, proofreading)
for deciphering a connectome becomes a major time bottleneck.
Here we introduce a large-scale, high-throughput, and semi-automated methodology to efficiently identify synapses.  We successfully applied our methodology to the Drosophila medulla optic lobe, annotating many more synapses than previous connectome efforts.  Our approaches are extensible and will make
the often complicated process of synapse identification accessible to a wider-community of potential proofreaders.
\end{abstract}

\section{Introduction}
Progress in the field of connectomics has led to large-scale reconstructions of the mouse retina \cite{Winfried13} and Drosophila optic lobe \cite{Nature13} using high-resolution, EM data.  A connectome consists of neurons and their connections, or synapses.  To discern neural connectivity, the authors in \cite{Winfried13} find putative synaptic sites between neurons through surface contact and a companion dataset.  Such a ``contactome'' \cite{Plaza14} can provide insufficient detail for understanding a connectivity graph and can produce misleading results as indicated in \cite{Nature13}.  Therefore, the authors in \cite{Nature13} determine connectivity by explicitly staining and manually annotating an EM dataset.  In these efforts, tracing small neuronal processes and verifying cell morphology were so time consuming that manually identifying the chemical synapses by exhaustively scanning the dataset was insignificant overhead.
But recent efforts in automatic image segmentation \cite{Parag14, Huang14, Andres, Funke12} to extract cell shapes promise a future where increasingly large connectomes are possible.  The minimal attention to automate synapse annotation will stifle this quest to uncover larger circuits \cite{Plaza14}.  Even a relatively small Drosophila brain contains tens of millions of synapses, which would would require years of annotation.

\begin{figure}
\centering
\includegraphics[width=1.0\textwidth]{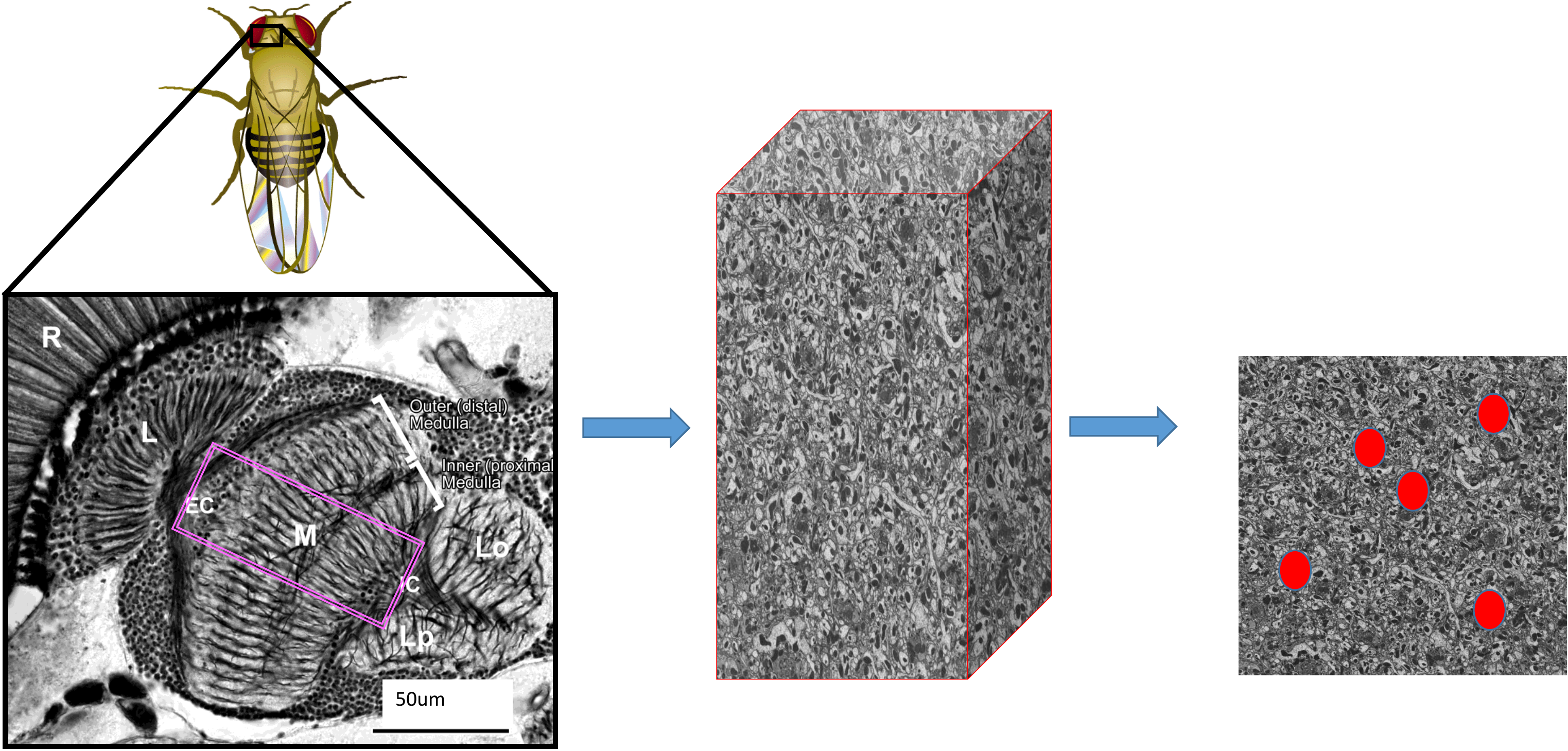}
\caption{\label{fig:highlevelflow} {\bf High-level flow for identifying synapses needed for a connectome.}  The rectangle in the left image outlines our region of interest -- seven columns of the Drosophila medulla.  We then image this region and annotate its synapses as indicated by the circles.}
\vspace{-1mm}
\end{figure}

\begin{figure}
\centering
\includegraphics[width=0.6\textwidth]{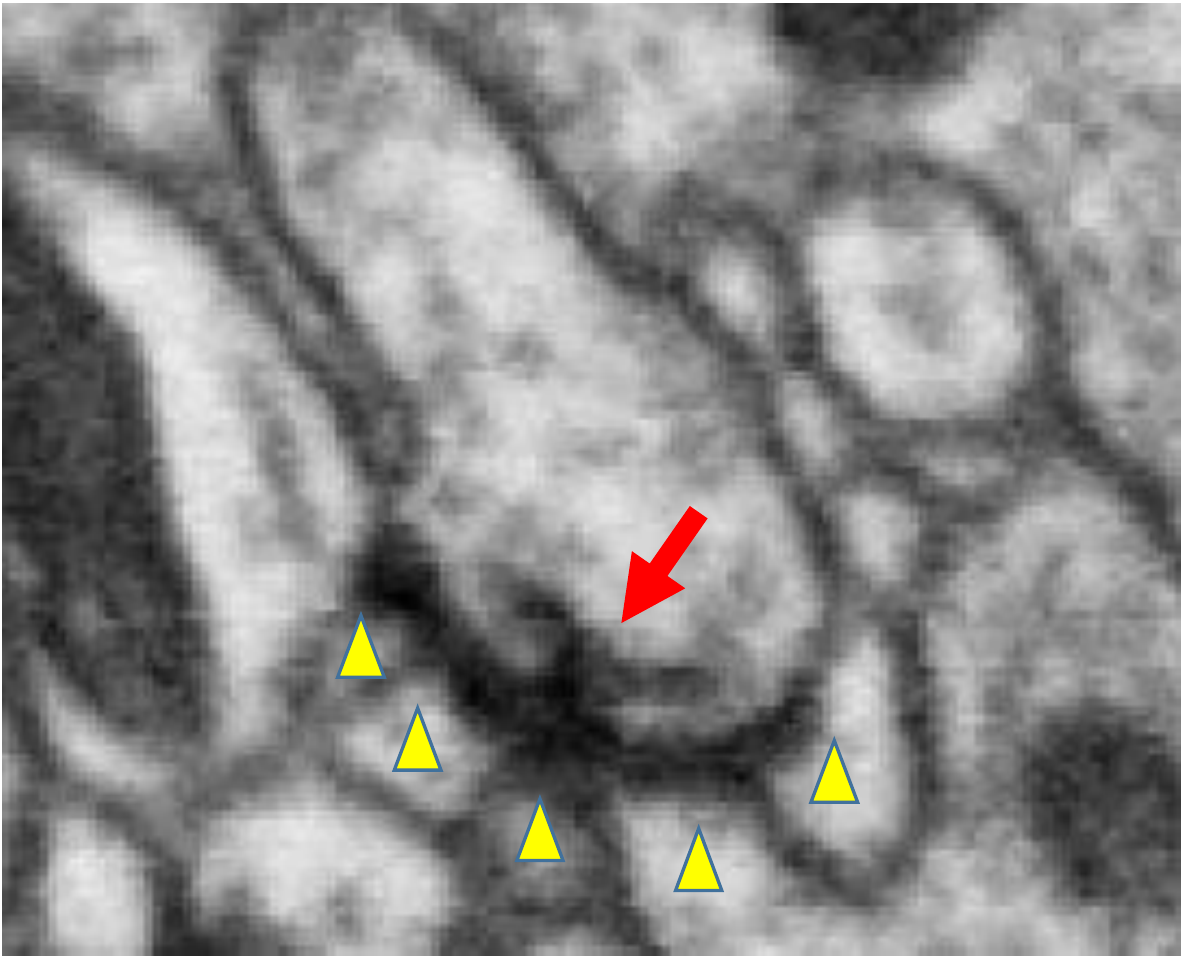}
\caption{\label{fig:synexample} {\bf Example of a synaptic site in the Drosophila.}  The arrow points to the pre-synaptic site called a T-bar.  Multiple outputs, called PSDs, exist, indicated by the triangles.}
\vspace{-1mm}
\end{figure}

This paper explores methodology to annotate synapses in large EM 
datasets to relieve bottlenecks in connectome reconstruction.  In particular, we annotate all the synapses in a region of the optic lobe of the Drosophila (seven medulla columns), shown in Figure~\ref{fig:highlevelflow}, as part of a larger effort to derive a connectome.  To annotate all the synapses in our dataset (40x40x50 µm, 10 nm per  voxel), we developed an algorithm to automatically identify putative 
pre-synaptic densities (called T-bars) in EM images.  We also 
introduced a workflow to manually validate T-bar predictions, and then 
identify the post-synaptic partners (called post-synaptic densities, or PSDs) for each T-bar, noting that 
Drosophila synapses are polyads, having more than one post-synaptic 
partner per T-bar as seen in Figure \ref{fig:synexample}.  The resulting annotations are used as information to guide subsequent
image segmentation and neuron identification. 
With our 
workflow, 4-6 proofreaders achieved a dense annotation of our dataset of 
unprecedented scale (56,621 T-bars and 336,735 post-synaptic partners) 
in 6 months.  This dense annotation greatly extends the previous state-of-the-art analysis of the one-column medulla annotation in \cite{Nature13}.
We observe high accuracy across proofreaders and good 
correspondence with published data, suggesting that these annotations 
reflect the actual distribution of chemical synapses.

Our work was greatly enhanced by our choice
of using FIB-SEM imaging to create the dataset \cite{fib}.
The near
isotropic resolution of the dataset enables superior image segmentation, faster analysis, and better opportunities for visualization.  Compared to the relative thick sections in
serial-section TEM, many more synapses were clearly identifiable.

Our methodology provides numerous contributions to both the EM reconstruction and computational communities:

\begin{enumerate}
\item Introducing a general workflow that allows for scalable and consistent annotations
\item Synapse prediction with accuracy comparable to state-of-the-art \cite{kresuk}
\item Semi-automated synapse identification that uses predictions to focus proofreaders' efforts
\item Visualization and focused workflows that reduce the complexity of analyzing complicated synapses, as seen in Figure \ref{fig:synexample}
\item Image segmentation to enhance synapse annotation
\item Algorithms that use synapse annotations to refine and guide subsequent image segmentation for reconstruction
\end{enumerate}

While this methodology is currently executed by a staff of trained 
editors, it makes the task more accessible to a wider audience and is 
scalable.  Moreover, we believe that further advances
in synapse prediction and visualization will make possible large-scale,
crowd-sourcing efforts.

In this paper, we first propose the general workflow.   Next, we 
introduce an efficient strategy to detect synapses
leveraging the interactive segmentation tool Ilastik \cite{ilastik}.  We then define software enhancements and protocols to enable faster synapse identification.  Finally, we provide results that breakdown our contributions and summarize the overall methodology.

\section{Workflow} \label{sec:workflow}
The assumed input to the workflow introduced here is an EM
image dataset.  In principle, the volume can be produced from 
serial-section TEM imaging, but more synapses will be easily identified if section thickness is small enough to produce near-isotropic
data, as with FIB-SEM imaging \cite{fib}.  The task of annotating
synapses requires markers (points) for each
pre-synaptic (T-bar) site and each
post-synaptic partner (PSD).  Due to the polyad arrangement of synapses
in Drosophila, we generally consider T-bar and PSD annotations
as separate work tasks.

In \cite{Nature13}, T-bars and PSDs were identified by the most
experienced proofreaders, which involved manually scanning the large image
dataset.  The challenge is twofold: 1) avoiding attentional errors
that might result in missing a site, and 2) deciding whether
ambiguous smudges or staining indicate T-bar, PSD, or neither.
We also note that the anisotropic dataset (3nm x 3nm x45nm) can
make identifying a T-bar, as shown in Figure \ref{fig:synexample}, very
challenging, depending on how the image plane cuts through
the T-bar.

\begin{figure}
\centering
\includegraphics[width=1.0\textwidth]{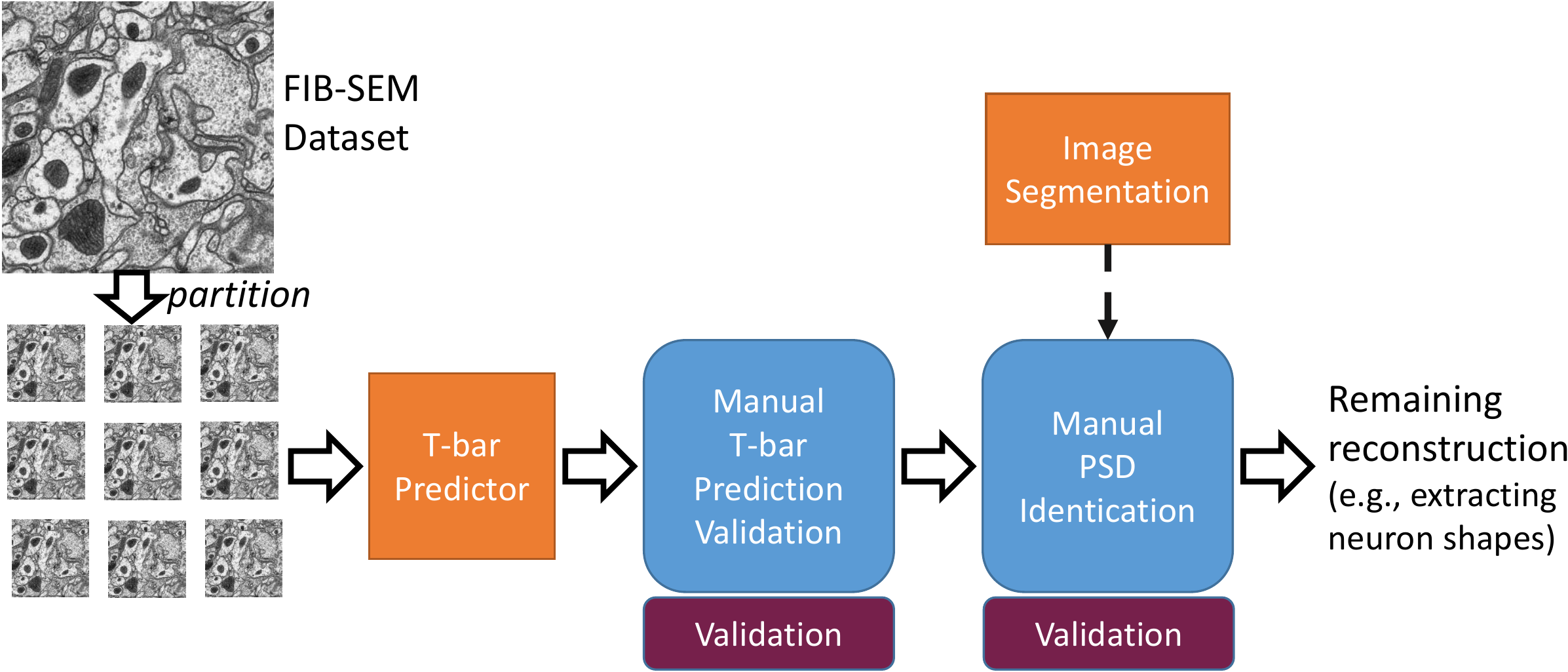}
\caption{\label{fig:synflow} {\bf Synapse annotation methodology.}  This flow uses automatic T-bar prediction
and image segmentation to boost the speed and consistency of manual annotation.  The annotations can then be used to enhance downstream extraction of neuron shapes needed when reconstructing a connectome.}
\vspace{-1mm}
\end{figure}

To improve upon this effort, we define a scalable annotation methodology
as outlined in Figure \ref{fig:synflow}.  We first collect FIB-SEM
images to eliminate the need for a dominant viewing plane when
annotating.  Then, we identify putative synapses through a T-bar
predictor introduced in Section \ref{sec:tbar}.  This predictor is trained on the pre-synaptic sites, as their visual cues are
stronger than those for PSDs.  We design the predictor to have high recall.  Proofreaders then examine
these sites, removing the predictions that were incorrect.  A subsequent
pass is done to identify the PSDs, using automatic image segmentation as a guide.

By decomposing the task of synapse identification into two passes, we can double-check our T-bars.  We do this by ensuring that the T-bar and PSD passes are assigned to different proofreaders.  We then stipulate that the proofreader checking the PSDs also verifies the corresponding T-bar annotations.  
The task decomposition into two passes also makes each pass more manageable, as the proofreader can focus on one type of action.  The flow for identifying PSDs is discussed more in Section \ref{sec:protocols}.

To implement this flow on a large dataset, we divide the volume into
several disjoint subvolumes.  The subvolume sizes are chosen to efficiently render in our visualization tool Raveler \cite{raveler} and to be a manageable unit of work for a proofreader.
Since each assignment
is disjoint and small, it would be easy to distribute widely.  As seen
in Figure \ref{fig:synflow}, we occasionally run validation stacks,
in order to assess the consistency of our proofreaders.

The above workflow attempts to reduce attentional recall errors by providing a predictor with high recall.  The possibility of a proofreader erroneously glancing over an image region as unimportant is minimized, since the proofreader need only examine the subset of locations determined to be likely T-bars by the automatic detection algorithm.  Furthermore, the double checking of synapses, use of image segmentation, and enhanced visuals and protocols described in Section \ref{sec:protocols} all aim to reduce the likelihood of errors and improve speed.

\vspace{2mm}
\noindent
{\bf Potential Application: Synapse-driven Segmentation}
\vspace{1mm}

While discussing the subsequent connectome reconstruction workflow (neuron extraction) is beyond
the scope of this paper, we introduce the idea of using the synapse
annotations to improve the quality of image segmentation (such as the segmentation produced in \cite{Parag14}).  This improvement is achieved by explicitly adding
constraints in the segmentation algorithms to avoid linking annotations
belonging to the same synapse.  In other words, all of the annotations
in Figure \ref{fig:synexample} would be guaranteed to be in different
image segments.  The intuition is that autapses and convergent PSDs from the same body are comparably rare.  By ensuring a conservative over-segmentation, tedious segment-splitting operations in segmentation-driven
reconstruction approaches \cite{Nature13}
can be avoided.  Furthermore, it is easy
to revisit disconnected T-bar and PSD segments when revising segmentation.  While our segmentation generally performs well in these synaptic regions independent of these constraints, some errors are made, as shown in the results, motivating the use of these constraints.  Therefore, we advocate
synapse identification before reconstructing neuron shapes, though segmentation can be used to aid both.

\section{Automatic T-bar/Synapse Detection} \label{sec:tbar}
Ideally, an automatic T-bar detector optimizes to near 100 percent
precision and recall.  However, many synapses are not very clear
in the image data, due to many factors such as imaging and preparation artifacts or the plasticity of the connection.  Often, human
interpretation is required to handle these ambiguities.  From a 
machine learning viewpoint, some of the features required to predict a 
synapse require more context than is readily exploitable or 
computationally feasible. 

As the predictions made by the T-bar detection will be verified by a proofreader, the misclassification cost associated with not predicting a T-bar is much higher than incorrectly predicting a T-bar where one does not exist, since the latter will be corrected by the proofreader but the former will not.  Therefore, our detector is targeted toward high recall, even if this means some sacrifice in precision.    
Subsequent improvements
to prediction can be quantified by accuracy or proofreading time, since higher precision equates to less work.

Here are the main aspects of our prediction algorithm:

\begin{enumerate}
\item Sparse voxel training for T-bar/not T-bar using Ilastik \cite{ilastik}
\item Application of voxel training to classify an EM image volume
\item Aggregation of connected voxel regions classified as T-bar
\item Clustering of similar regions to improve algorithm precision
\end{enumerate}

\subsection{Voxel Training and Prediction}

Our objective is to automatically detect the voxel locations of pre-synaptic densities given an EM image volume. To achieve this, we have trained a voxel classifier that should classify a voxel as 1 if it is part of a synaptic density and 0 otherwise. Since the data is noisy, our choice of classifier was Random Forests (RF) due to its robustness against noise and its parallel-training capability. The RF predictor assigns a real-valued confidence for each voxel, which is transformed into a binary decision by thresholding.

\begin{figure}
\centering
\includegraphics[width=1.0\textwidth]{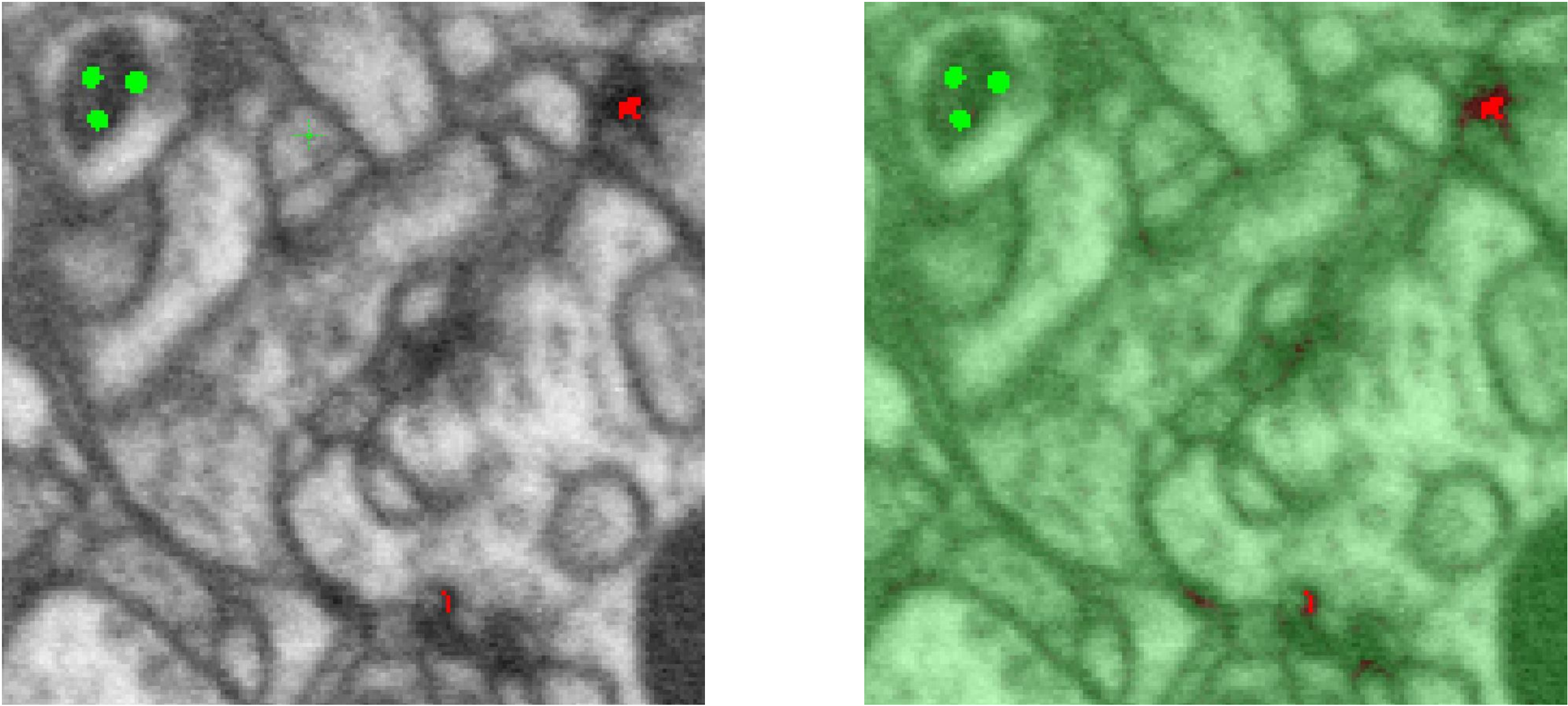}
\caption{\label{fig:training} {\bf Training a voxel classifier interactively in Ilastik.} The left picture shows manual labels for synapses and background.  The right picture shows the resulting voxel-level prediction.}
\vspace{-1mm}
\end{figure}

To perform the RF training, we use the interactive tool Ilastik \cite{ilastik}.
Figure \ref{fig:training} shows one slice of a grayscale image with a few training points.  
While the amount of training data needed is often small, we note that the quality of the training labels can have immense ramifications on the quality of the
algorithm.  In particular, we note that the best results require careful negative-class labeling of non-T-bar densities, such as mitochondria, whose appearance is similar to T-bars.
Also, one needs to iterate many times with the Ilastik tool and
evaluate the downstream, post-processed results to understand
the ramification of label decisions.  In general, it is better to more aggressively predict synapses and let post-processing remove false positives.

\subsection{Post-processing}

The resulting prediction produces connected components of voxels labeled as 1 (belonging to a T-bar).  We then 
cluster them into
components to produce synapse prediction locations.

To cluster these regions, we apply a weighted or Fuzzy K-means algorithm. The real-valued confidences of the RF classifier act as weights in our formulation. More precisely, clustering
is achieved by minimizing
weighted distances from the $K$ centers $\mathbf{c}_k$ to each of the voxels
$i$ in the connected components:
\begin{equation}
\sum_k^K \sum_{x_i \in k} p(\mathbf{x}_i)~|| \mathbf{x}_i - \mathbf{c}_k||^2
\end{equation}

where $p(\mathbf{x}_i)$ is the RF output at $\mathbf{x_i}$.
Effectively, our formulation is biased to converge near concentrations of high confidence values. A non-maxima suppression removes multiple centers in close proximity from the output of weighted K-means.  If the algorithm incorrectly
removes a center, a proofreader could still annotate
the missing T-bar since the proximity will be close to another
prediction. 

The efficiency of the post-processing approach enables rapid
feedback to the person performing training.  The entire flow
is computationally non-intensive and parallelizable
as the features
used in voxel prediction are local.

\section{Focused Annotation Protocols} \label{sec:protocols}
To best exploit the T-bar predictions and implement
the flow introduced in Figure \ref{fig:synflow}, we devise
a series of visual and workflow enhancements to the publicly available tool Raveler \cite{raveler}.  In particular, we implement very specific protocols for T-bar and PSD identification.  Given a list of T-bar predictions, Raveler
will create a sequence of sub-tasks that allows a proofreader to efficiently process and focus on one prediction.  Then, given
verified T-bars, Raveler will create a sequence of
sub-tasks that allow a proofreader to annotate its post-synaptic partners.  We
define these flows as {\em focused annotation}.

\begin{figure}
\centering
\includegraphics[width=0.5\textwidth]{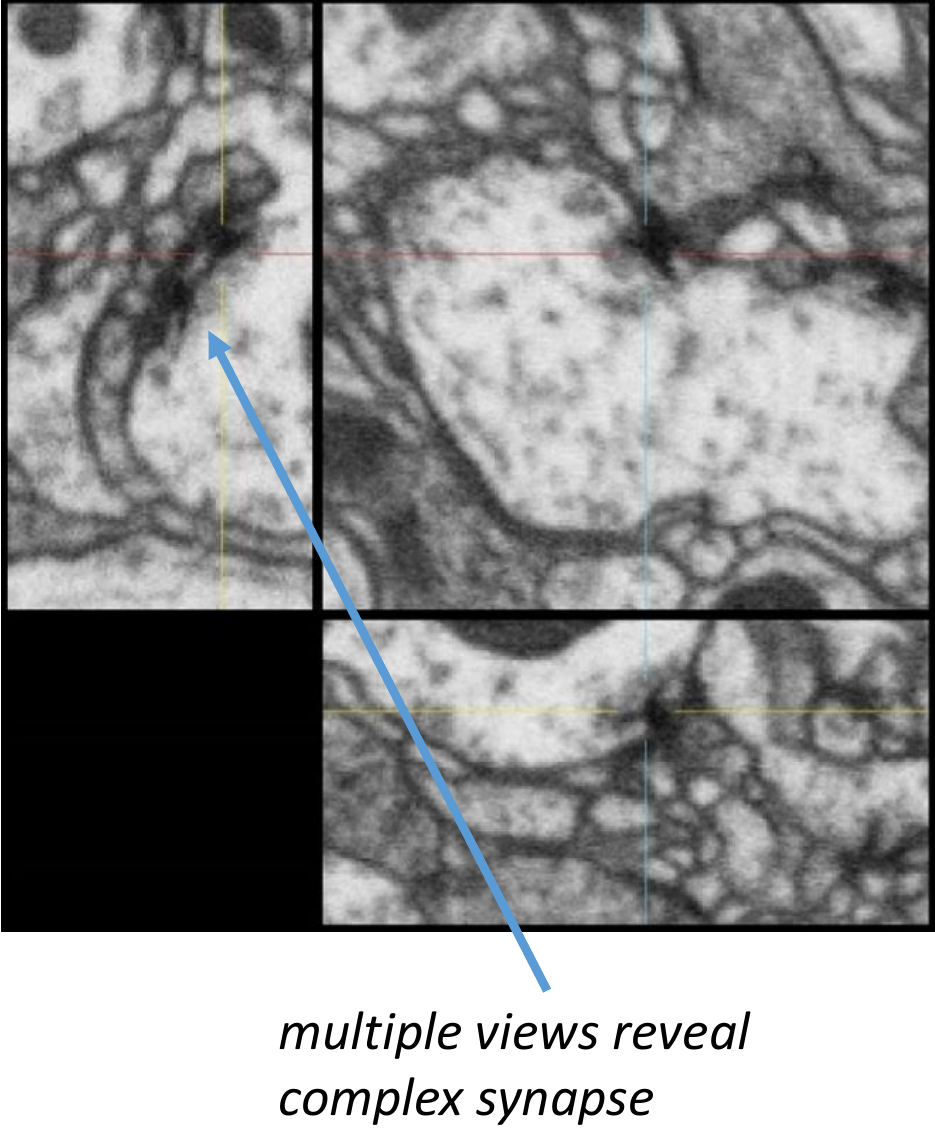}
\caption{\label{fig:tbarorthog} {\bf Orthogonal viewing planes to visualize complex synapse shapes.}  Features of a synapse might not be visible in just one plane of viewing.  This figure show a multiple pedestal T-bar (complex synapse), only visible in one of the views.}
\vspace{-1mm}
\end{figure}

The T-bar 
validation is more involved than a simple yes/no question, as we allow the 
proofreader to choose a more ideal location for the annotation.  As a consequence of this freedom, the proofreader will sometimes identify multiple T-bars or another nearby T-bar for a given prediction, effectively increasing recall.

We next highlight a few visualization enhancements for focused annotation and the use of segmentation to guide PSD annotation.

\subsection{Visualization Enhancements}
We exploit isotropic image data by implementing 3D
orthogonal viewing, as shown in Figure \ref{fig:tbarorthog}.
It is often the case that a single image plane misses
or incorrectly characterizes a particular biological
structure.  In Figure \ref{fig:tbarorthog}, we see the
presence of a {\em complex synapse} (a synapse with multiple
T-bars or vesicle release sites) \cite{complex1, complex2}
in only one of the image planes.  Our
protocols will automatically open such a viewer to
the predicted T-bar location, making it more likely
that a proofreader will discover it.  Because the context
around a synapse does not need to be very large to identify
T-bar and PSD elements, we can potentially perform additional visualization optimization.  For T-bar verification, this further emphasizes the need for a predictor that also gives optimal locations.

We also carefully created a set of glyphs (or markers)
in Raveler for T-bars and PSDs (with additional considerations for sub-categories like complex T-bars).  While this may seem trivial, the density of annotations make it easy for clumsily designed glyphs to overlap or otherwise make it difficult to discern the given annotations.  Furthermore, we must carefully show annotations/glyphs that were placed on different, but nearby, viewing planes, so that proofreaders can avoid doubly annotating the same synapse.  The protocol also helps prevent this by adding simple distance constraints.

\subsection{PSD Workflow}\label{sec:psd_workflow}
Identifying PSD sites is much more challenging and time-consuming than verifying T-bars.  PSD sites for individual neurons are often
very small and also may require tracing through regions only 15-20nm thick.  Furthermore, the PSD connections will form at many angles to the T-bar region.  

To ameliorate the problem, one approach, planned for future work, is to 
identify PSDs automatically given a T-bar prediction, perhaps using automatic image segmentation and T-bar proximity.  Then, we could provide a list of
points for proofreaders to verify.
This strategy has two obstacles: 1) image segmentation still produces a 
lot of errors, presumably more in very fine neuronal regions, and 2) neurons that are close to a synaptic zone might not synapse.  Often, a lot of human expertise goes into deciding
whether two neurons have enough evidence to form a connection.  

\begin{figure}
\centering
\includegraphics[width=1.0\textwidth]{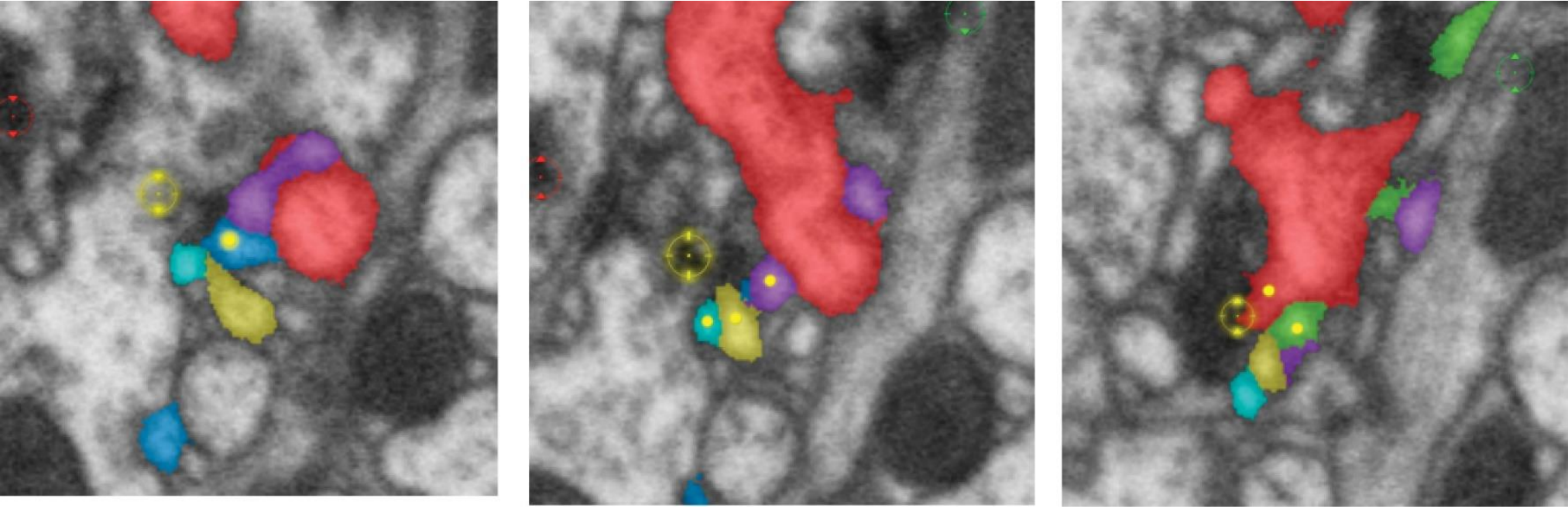}
\caption{\label{fig:psdflow} {\bf PSD annotation methodology.}  Segmentation acts as a guide to annotations, allowing proofreaders
to prune examined areas visually.}
\vspace{-1mm}
\end{figure}

While it may be feasible to identify PSDs using some probabilistic model in the future, at the moment we simply
use automatic segmentation as a guide for manual proofreading.
Figure \ref{fig:psdflow} shows an annotated T-bar and surrounding segmented bodies.  When the proofreader gets
the assignment, none of the bodies are highlighted.  As
he/she marks PSD sites, the underlying segmentation
is highlighted.  This effectively prunes the visual space.
The proofreader can easily detect, by scrolling through
a single planar
view, if all adjacent regions have been clicked.  (While some adjacent regions do not have PSDs, most do).  It will
also limit scenarios where a proofreader double marks a
PSD on the same body.

While image segmentation is not perfect in these synaptic
regions, it is not required to be.  The segmentation primarily must avoid under-segmentation, so that multiple
neurons are not pruned with a single annotation.  To
address this, we
use segmentation produced by \cite{Parag14}, which favors
over-segmentation.  We report on the quality of this
segmentation in the experiments.  Over-segmentation is less
problematic, but will mean that pruning will eliminate fewer segments and double annotation will be more likely.  However, double annotation can again be partially minimized by considering
the proximity of annotations.

\section{Experiments}
We evaluate our algorithms and workflow on an EM dataset
with 10nm voxel resolution produced by FIB-SEM imaging \cite{fib}.  The dataset
is $27,000$ cubic microns and constitutes 7 columns of medulla
in the Drosophila optic lobe.  For computational efficiency
and to create manageable assignments for proofreaders,
we divide this volume into 125 cubic micron subvolumes.  All annotation
was done in Raveler \cite{raveler} by 4-6 trained proofreaders.

\begin{figure}[htb]
\centering
\includegraphics[width=0.8\textwidth]{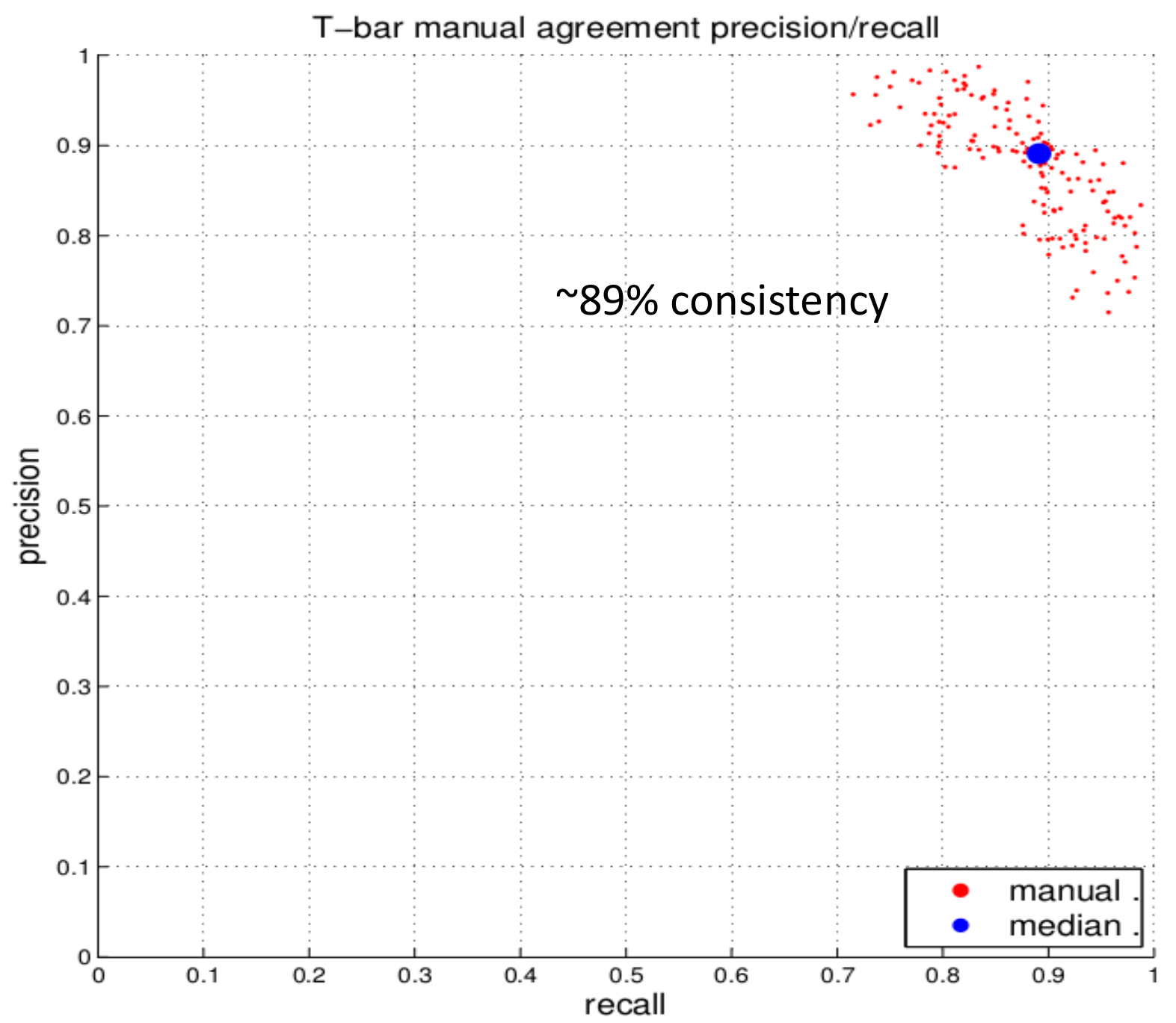}
\caption{\label{fig:proofconsistency} {\bf Consistency of synapse annotations among proofreaders.}  This plot shows high consistency
between a set of proofreaders in production stack.  The accuracy gives an estimate for how accurate T-bar prediction should be to allow for complete automation (no proofreading).}
\vspace{-1mm}
\end{figure}

\subsection{Evaluation of T-bar Prediction}
To assess the quality between two sets of T-bar annotations,
we generate a precision-recall curve by 
an optimal matching
algorithm with a distance constraint (matches are not possible
beyond a threshold of 30 voxels/300 nm, roughly the width of a T-bar, in these experiments).
More precisely, we formulate a binary integer programming problem
where Euclidean distance defines the cost of a potential match (between predicted and ground-truth T-bar locations) and distances beyond
$30$ voxels will not reduce total cost.  Precision-recall curves
can then be computed by varying the threshold applied to
the initial voxel
prediction produced by Ilastik \cite{ilastik}, with all other parameters fixed.


We first show the consistency of proofreaders using our T-bar protocol
over several subvolumes in Figure \ref{fig:proofconsistency}.
Note that high variance between proofreaders could indicate the
need to have multiple annotators per subvolume or that the problem is ill-posed at the given image resolution and noise.
Fortunately, the plot
indicates very high agreement between all pairs of proofreaders, with consistency close to 90\%, and suggests that having just one proofreader assigned to each subvolume is reasonable.

A small percentage of subvolumes were annotated by multiple proofreaders, for the purposes of quality control and measurement of manual agreement as reported above.  For the remaining majority of subvolumes, a single proofreader reviewed the annotations produced by the automated T-bar detector, tuned to high recall (90\%, mirroring human consistency).  The reviewed annotations then form our ``ground-truth''.


\begin{figure}[htb]
\centering
\includegraphics[width=1.0\textwidth]{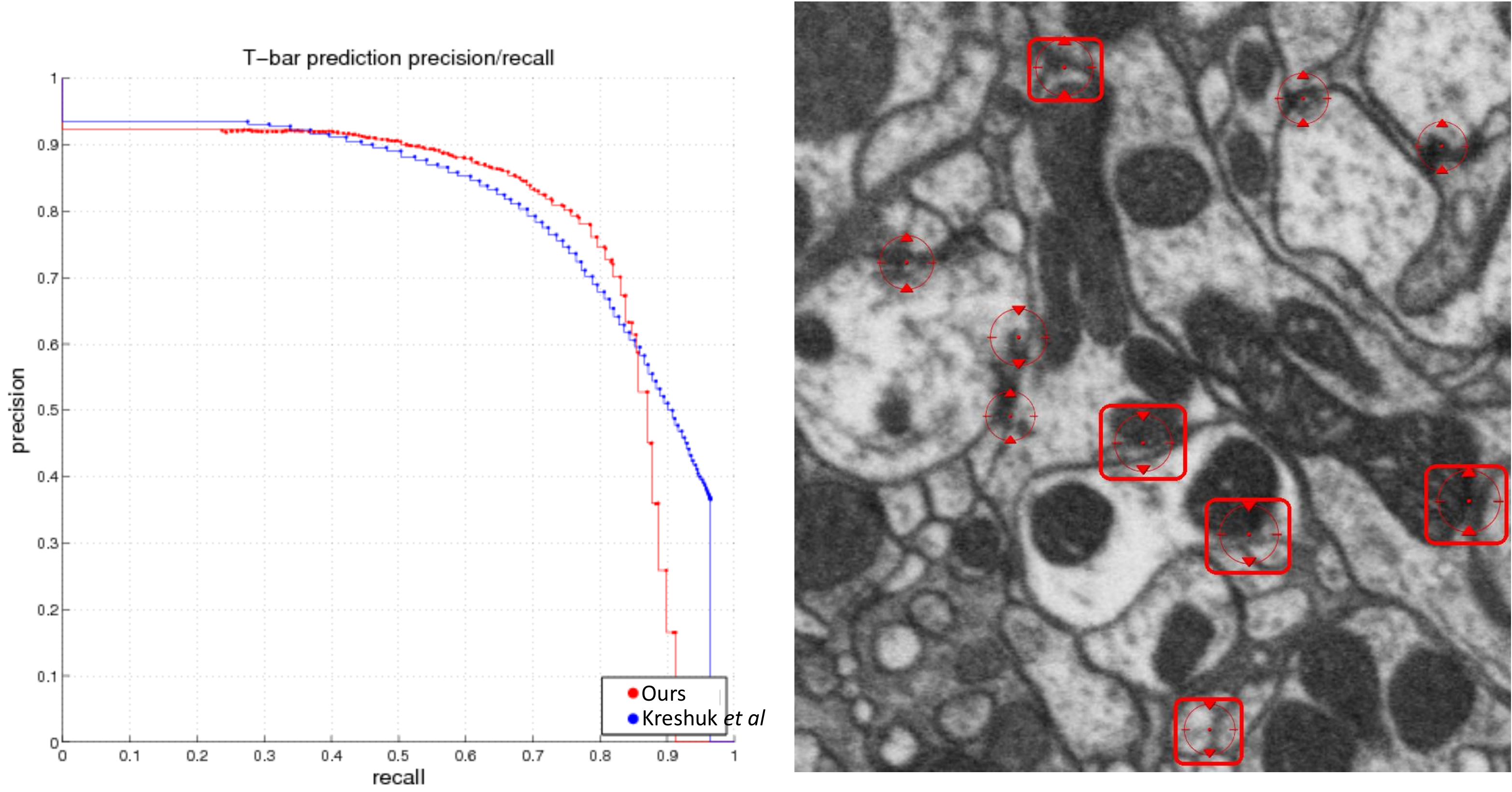}
\caption{\label{fig:predresults} {\bf Quality of synapse prediction.}  Left) Despite our simple T-bar prediction approach, we achieve around 50\% precision given 90\% recall.  This compares
favorably to other, more complex approaches \cite{kresuk}.  Comparisons are made over a set of 20 subvolumes.  Right) Examples of predictions (in circles); false positives are marked with squares.}
\vspace{-1mm}
\end{figure}

The left part of Figure \ref{fig:predresults} shows the accuracy of 
our prediction
compared to the ground-truth over several subvolumes.  We achieve
similar performance compared to the work in \cite{kresuk}.  We
believe this performance is due to the
care in placing Ilastik training labels.  Note that the recall
is not 100\%.  As mentioned before, proofreaders often annotate
nearby unpredicted T-bars and may search locally around
the prediction if the predicted location is inaccurate or spurious.
(Ideally, we desire to optimize the location of the T-bar prediction, as well, since a better location reduces searching by the
proofreader.)


Qualitatively, at our operating parameter values
(of high recall), we observe many false positives when image
features resemble a T-bar, such as on or near the
mitochondria bodies and where mitochondria are close
to the membrane.
The right part of Figure \ref{fig:predresults} shows 
false positives enclosed by rectangles.
Unfortunately, training a mitochondria voxel detector and using it
as a feature for the synapse voxel predictor did not improve
performance.  This may be due to voxel
features that are not sufficiently descriptive to
achieve a higher accuracy.  A better solution might involve
a patch-based detection approach where a small patch or window
is classified as a single unit.


\begin{table}[htb]
\centering
\begin{tabular}{|l|r|r|r|r|} \hline
Proofreader & \multicolumn{2}{c|}{hours} & \multicolumn{2}{c|}{$\#$ T-bars} \\ \hline
 & unguided & guided & unguided & guided \\ \hline \hline
A & 6.08  & 6.44  & 960  & 873 \\
B & 12.44 & 9.03  & 1028 & 888 \\
C & 19.44 & 11.63 & 927  & 717 \\ \hline
\end{tabular}
\caption{
\label{tab:manual_tbar_comp}
{\bf Comparison of unguided and guided workflows for T-bar detection.}  For each proofreader, the total amount of time spent to annotate four subvolumes, as well as total number of T-bars annotated, is given, for both workflows.  The difference in total annotation time between the fastest and slowest proofreader is smaller when using the guided approach.}
\end{table}

Additionally, we have run experiments comparing our T-bar workflow, which uses automated T-bar detection guidance, versus a completely manual, unguided approach.  Four subvolumes were randomly selected, and each subvolume was manually annotated by each of three proofreaders, first under an unguided approach without any automated detections, and second with our automated, guided workflow.  The amount of time taken by each proofreader, as well as the number of detected T-bars, is given in Table~\ref{tab:manual_tbar_comp}.

We observed that proofreaders consistently detected less T-bars when using the guided workflow; over the four subvolumes and three proofreaders, the number of detected T-bars using the guided workflow was 85\% of the number of detected T-bars using the unguided workflow.  This variability is slightly greater than the consistency across proofreaders of about 90\%, and suggests some amount of missed T-bars when relying on the automated detections.  However, the total time spent using the guided workflow was 71\% of the time spent using the unguided workflow.  This saving in time was more pronounced for proofreaders who required more time under the unguided workflow.

Although the recall for manual annotation is slightly better using highly-trained proofreaders, we speculate that our guided T-bar workflow is particularly advantageous for less experienced proofreaders.  As our proofreaders are both experienced in general, as well as experienced with this particular data set, their unguided annotation results are in some respects an upper bound on expected performance without automated guidance.  For untrained or inexperienced proofreaders, we would expect an unguided approach would take longer and be less consistent, and therefore see a larger improvement from a guided workflow than observed in our experiment.  Lastly, further optimizations in machine predictions should reduce the number of missed T-bars and potentially further decrease the amount of time required for manual verification.


\subsection{Evaluation of PSD Workflow}
We enhanced the protocol to identify PSDs by providing segmentation guidance, as discussed in Section~\ref{sec:psd_workflow}.  To assess the viability of using segmentation
as a guide, we evaluate the number of times a segmentation
violates a synapse constraint in a ground-truth 125 cubic micron
volume.  We consider two synapse constraints: 1) a segment cannot
connect a T-bar and partner PSD, 2) a segment cannot connect two
neighboring PSDs.  

Given a ground-truth synapse annotation and predicted segmentation, we refer to any violation of the above constraints as an {\em under-segmentation violation}.  When identifying PSDs using segmentation guidance, an under-segmentation violation implies that a proofreader may potentially miss an annotation, for instance, because two ground-truth PSDs exist within the same predicted segment.  Note, however, that an under-segmentation violation is not necessarily a {\em biological violation}, as it may be the case that the ground-truth data itself violates the constraint (for example, an autapse violates the first constraint).  Table \ref{tab:overseg} indicates a very small number of under-segmentation violations, only $13$ of which are biological violations where a predicted segment incorrectly merged two ground-truth bodies.  Only 5\% of all synapses (T-bars plus PSDs) were impacted by these biological violations.


In Section \ref{sec:workflow}, we introduced the potential
to use synapse annotations to guide subsequent automatic image
segmentation.  We note that while proofreaders encounter under-segmentation violations in the PSD protocol, the segmentation is only used as a guide and a proofreader can and will ignore under-segmentation errors.  Because of this, we can apply these annotations to create constraints for subsequent segmentation, so that the under-segmentation errors are not repeated.  For this example, we note that the ground-truth volume (synapse annotations and segmentation) only has $43$ under-segmentation violations out of $1957$ possible pairwise constraints.  Therefore, the synapse annotation constraints will not adversely result in over-segmentation.

\begin{table}[htb]
\centering
\begin{tabular}{|l|l|}\hline
under-segmentation violations & 37  \\
biological violations & 13 \\
total synapses (T-bars) & 260 \\
\% synapses w/biological violations & 5\% \\
\hline \end{tabular}
\caption{
\label{tab:overseg}
{\bf Evaluation of under-segmentation errors around synapses.}  We denote under-segmentation violations as places where
T-bar and their PSDs are in the same segment.  These violations
will not always result in biological errors.  Biological
violations occur only $13$ times impacting
only 5\% of the synapses.}
\vspace{-1mm}
\end{table}

\begin{table}[htb]
\centering
\begin{tabular}{|l|r|r||r|}\hline
Proofreader & psds/hr w/o seg & psds/hr w/seg & improvement \\ \hline \hline
A & 77 & 101 & 32\%  \\
B & 132 & 208 & 58\% \\
C & 101 & 125 & 24\% \\
D & 282 & 308 & 9\% \\
E & 157 & 187 & 20\% \\
F & 107 & 136 & 27\% \\ \hline
all & 143 & 178 & 25\% \\
\hline \end{tabular}
\caption{
\label{tab:flow}
{\bf Comparison of PSD annotation
times with and without segmentation-aided visualization.}
Significant speed-ups occur for all
proofreaders when using segmentation.}
\vspace{-1mm}
\end{table}

The segmentation-guided workflow was actually implemented after
over half of the production annotation was complete.  Although not a
direct analysis of performance benefits, we are able to indirectly compare
the per PSD annotation rates
of individual proofreaders with and without segmentation across
different substacks.  Table \ref{tab:flow} shows the PSD
annotations per hour for three different subvolumes and for
six different proofreaders.  We observe consistent speed-ups across proofreaders.

\subsection{Evaluation of Entire Workflow}
We annotated the entire 7-column dataset, identifying 56,621 T-bars
and 336,735 PSDs.  The resulting density of T-bars per cubic micron
is 2.1, higher than the 1.4 reported for the one-column medulla
reconstruction in \cite{Nature13}.  While some of this difference is likely due to slightly different procedures concerning what should be annotated, analysis of
data suggests that the ability to observe T-bars in orthogonal
views in our isotropic FIB-SEM dataset
contributes to a higher absolute recall. The 40-50nm section
thickness in \cite{Nature13} made orthogonal viewing less practical.
Furthermore, we achieve a higher ratio of PSDs per Tbar of 5.9 than
reported in \cite{Nature13}.  Similarly, this higher ratio is likely a result
of being able to more easily identify small processes that form
synapses at angles oblique to the section plane.  We also suspect that
the focused protocols contributes to a more thorough result.
While true accuracy is hard to assess, the relative similarity to the distribution of synapses seen in \cite{Nature13}, coupled with the bounds of the T-bar prediction algorithm, inspires confidence.

\begin{figure}
\centering
\includegraphics[width=1.0\textwidth]{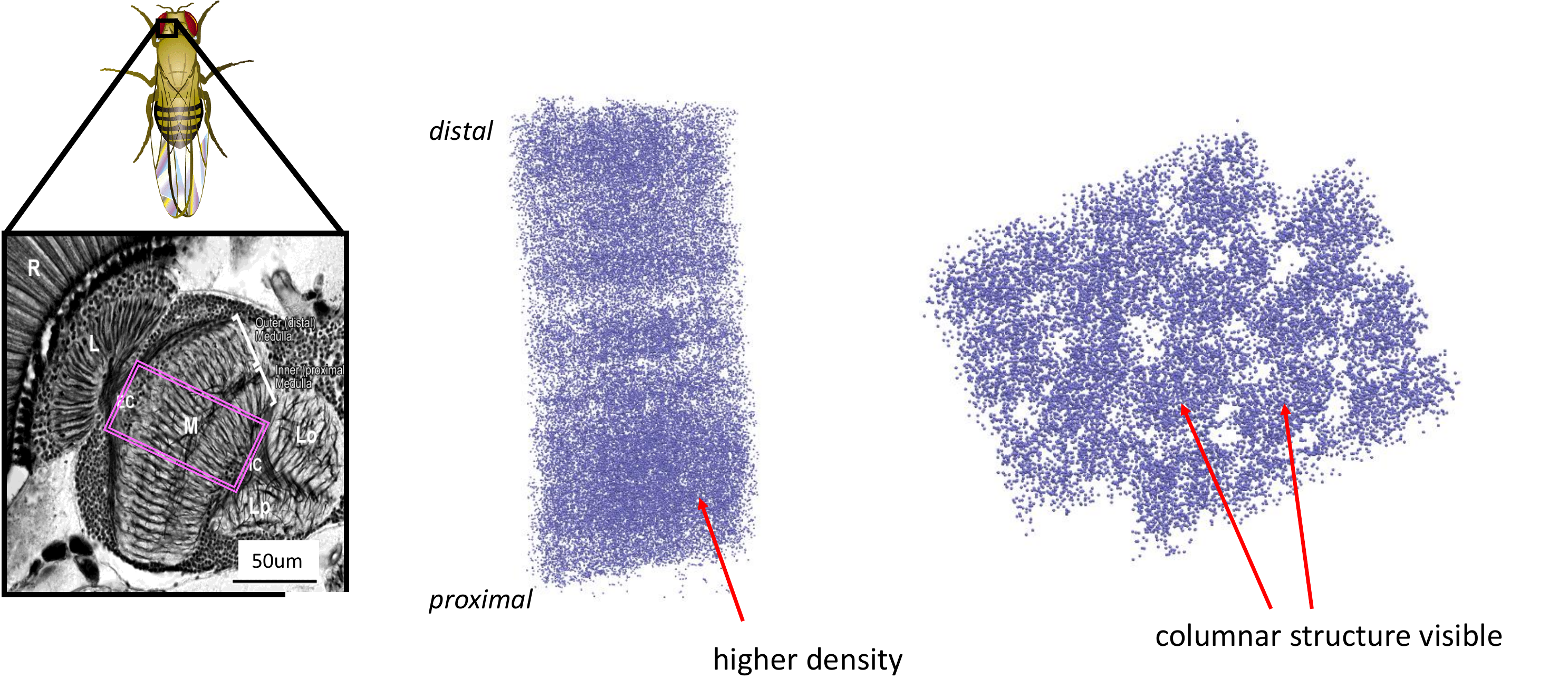}
\caption{\label{fig:synapsecloud} {\bf Illustration of all synapses in the dataset.}  Even from this visualization, one can discern that the proximal region of the medulla (the lower part of the picture in the center) contains a higher density of synapses.  The picture on the right, showing a different viewing projection, reveals the columnar structure of the underlying dataset.}
\vspace{-1mm}
\end{figure}

While this paper does not go into detail about biological findings,
we note that
our annotations allow for interesting high-level analysis.
Figure \ref{fig:synapsecloud} shows a point
cloud representation of every T-bar in the dataset.  The left cloud
shows an X/Z axis where the top is the distal medulla and the bottom
is the proximal.  One can notice the higher density in the proximal
region, indicative of increased inter-columnar activity.  In the X/Y
projection in the right cloud, one can discern different medulla columns,
thus indicating some stereotypy in organization between them.  Such
high-level analysis could be an effective mechanism to accurately
identify different brain compartments.

\begin{figure} 
\centering
\includegraphics[width=1.0\textwidth]{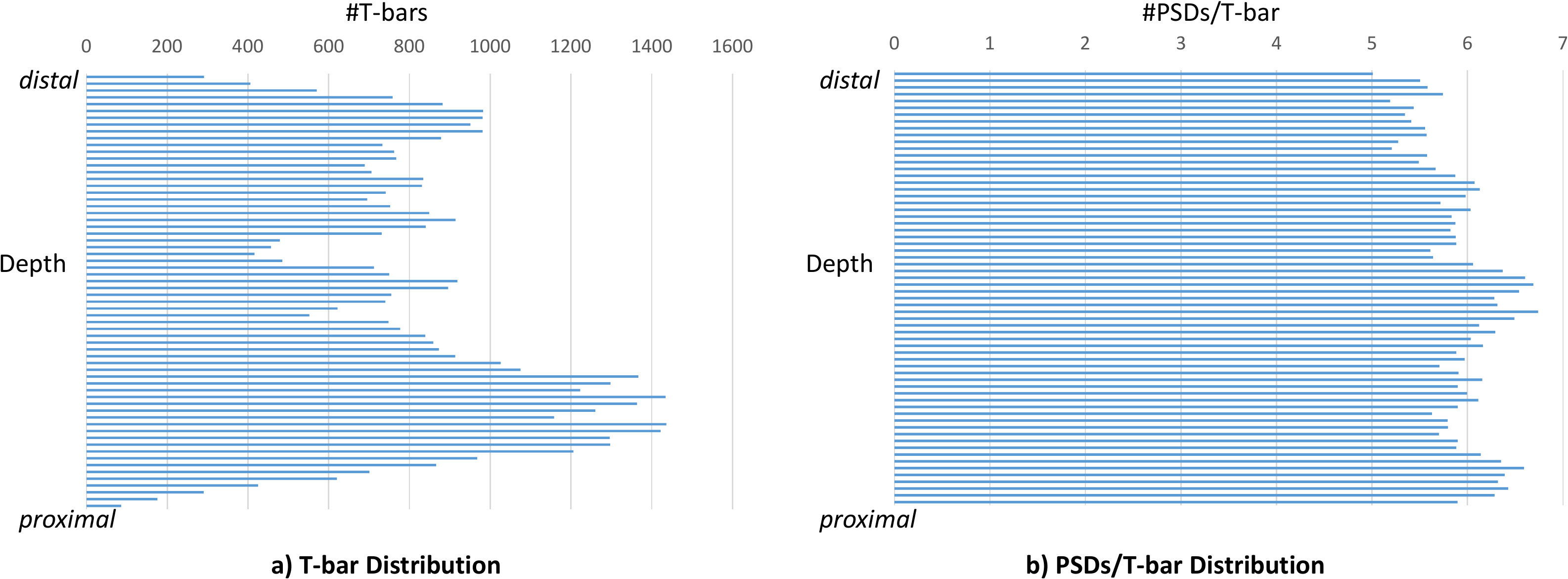}
\caption{\label{fig:zdist} a) Distribution of T-bars
as a function of column depth.  Maximum densities in the proximal layers are over twice those of the minimal densities.  b) Distribution of PSDs per T-Bar is more consistent.}
\vspace{-1mm}
\end{figure}

Figure \ref{fig:zdist} shows how T-bars and PSDs per T-bar vary in Z.  Each
bin represents 1 micron.
As seen in the point cloud, there is a higher density in the more proximal
layers.  Conversely, the PSD per T-bar ratio is much more consistent throughout.

Lastly, we report the average time it takes to annotate a subvolume.
{\tt session hrs} refers to the time it takes to complete a subvolume,
while {\tt working hrs} tries to account for the subset of time that is
actually clicking and navigating in Raveler, factoring out
inherent inefficiency such as when there are ambiguities and extra
thought is required.  Working hours to session hours give {\tt efficiency}, which could be seen as a {\em frustration} index.  Frustrating tasks often lead to less work efficiency.  Note that T-bar
annotation is not the time bottleneck in general.  Despite improvements
to both protocols, PSD tracing is time consuming and requires a lot of visual inspection of the data.  With our 4-6 person proofreader team, we completed the entire volume in 6 months.

\begin{table}
\centering
\begin{tabular}{|l|r|r|r|r|}\hline
Task & session hrs & working hrs & efficiency & microns/day \\ \hline \hline
tbar ann & 352 & 309 & 88\% & 613 \\
psd ann & 2582 & 1973 & 76\% & 84 \\
average & & & & {\bf 73.6} \\
\hline \end{tabular}
\caption{
\label{tab:7col}
{\bf Breakdown of synapse annotation effort in the Drosophila medulla.}
All of these tasks are performed on subvolumes 125 microns in size.}
\vspace{-1mm}
\end{table}

\section{Conclusions}
We introduce and deploy a semi-automatic, scalable
synapse annotation workflow on a large EM dataset, thoroughly identifying synapses in a specific region of unprecedented
volume.  Our results highlight the effectiveness of our synapse prediction and the positive
impact of focused annotation workflows in proofreader
productivity.  We also introduce the concept of
synapse-driven connectome
analysis.  The identification of synapses can reveal broad patterns of connectivity and be used as constraints for automatic segmentation in downstream
analysis.

The techniques proposed are amenable to crowd-sourcing.  Future methodology should leverage improvements to T-bar predictions, new PSD identification algorithms, and novel visualization strategies to achieve further speed-ups and
broaden accessibility.  
However, even with perfect prediction, laborious verification is required if all predicted sites
are examined.  Fortunately, in neuropil with many synapses between connecting neurons (for instance in the Drosophila), 
it may be possible to completely automate synapse annotation through a higher-precision, lower-recall
approach.

{\small
\textbf{Acknowledgements:} We thank Zhiyuan Lu for sample preparation, Shan Xu and Harald Hess for FIB-SEM imaging; Shin-ya Takemura, and
the FlyEM proofreading team (Roxanne Aniceto, Lei-Ann Chang, Shirley Lauchie, Christopher Sigmund, Satoko Takemura, Julie Tran) for biological guidance and the reconstruction efforts; Ting Zhao for visualizations; and Louis Scheffer and Ian Meinertzhagen for useful discussions and suggestions.}

\end{document}